\documentclass[a4paper,11pt]{article}
\pdfoutput=1 

\usepackage{jinstpub} 

\usepackage{graphicx}
\usepackage{rotating}
\usepackage{amssymb}
\usepackage[numbers]{natbib}
\usepackage{subfig}
\usepackage{float}
\usepackage{multirow}

\title{\boldmath Vibration decoupling system for massive bolometers in dry cryostats}

\author[a,1]{R. Maisonobe,\note{Corresponding author.}}
\author[a]{J. Billard,}
\author[a]{M. De Jesus,}
\author[a]{A. Juillard,}
\author[a]{D. Misiak,}
\author[b]{E. Olivieri,}
\author[a]{S. Sayah,}
\author[a]{and L. Vagneron}

\affiliation[a]{Univ Lyon, Universit\'e Lyon 1, CNRS/IN2P3, IPNL-Lyon, F-69622 Villeurbanne, France}
\affiliation[b]{CSNSM, Univ Paris-Sud, CNRS/IN2P3, Universit\'e Paris-Saclay, 91405 Orsay, France}

\emailAdd{r.maisonobe@ipnl.in2p3.fr}

\abstract{Pulse-tube based dilution refrigerators are massively employed in low temperature physics. They allow to reduce the running costs and to be operated with unprecedented easiness. However, the main drawback of this technology is the mechanical vibrations induced by the pulse-tube cryocooler. These perturbations can cause extra-noises drastically affecting the detector performance. In this paper, we propose a solution to mitigate the impact of these vibrations by mounting the detectors in an elastic-pendulum based suspended tower. Based on vibration modeling and experimental tests, we show that the vibration levels are attenuated by up to two orders of magnitude at most frequencies, especially above $\sim20$~Hz, for both vertical and radial directions. Thanks to this passive isolation solution, vibration levels, both along vertical and radial directions, below 1~$\upmu\textrm{g/}\sqrt{\text{Hz}}$ in the frequency range [1-1000]~Hz are obtained. This provides a convenient environment to test the ultimate performance of low temperature detectors. As a result, we report an improvement by one to two orders of magnitude on the noise levels of massive cryogenic bolometers, leading to thermal energy resolutions improved by a factor 5 to 40. Finally, we conclude that the energy resolution of our cryogenic bolometers are no longer limited from any residual vibrations, hence allowing the perspective of further improving our bolometer performance in the context of low-mass dark matter searches and neutrino physics applications.}


\arxivnumber{} 

\begin{document}
\maketitle
\flushbottom

\section{Introduction}
\label{sec:intro}
Dry Dilution Refrigerators (DDR) are a convenient alternative to Wet Dilution Refrigerators (WDR) in terms of low-cost and ease of operation. They have a high level of automation and avoid the consumption of liquid helium while providing a similar low temperature environment as the one obtained with WDR. The cool-down process of the two first stages within DDR is ensured by the technology of pulse-tube cryocoolers. However, the mechanical vibrations they induce can drastically affect the performance of cryogenic detectors. This effect has been reported by several cryogenic experiments using such bolometers \cite{key-1,key-2,key-3}.

Several approaches have been considered to mitigate vibrations. In the late 90s, the CUORICINO collaboration was the first to test the idea of a decoupling system to improve performance of cryogenic detectors. A first attempt was made to suspend a massive TeO\textsubscript{2} detector array with a Teflon damped steel spring \cite{Pirro_1}. The system was improved in \cite{Pirro_2} to a two-stage low-pass mechanical filter by adding an intermediate lead disk anchored to the cryostat by three stainless steel wires and harmonic stainless steel strips.

The CUORE experiment has implemented a different strategy. A Y-Beam (with three connecting points) at 300~K isolated from the cryostat through \emph{Minus-K} suspensions supports the whole 988 TeO2 detector array. Despite the use of three CRYOMECH PT415 pulse tubes, they report keV-scale detector energy resolutions \cite{cuore0,cuore}. 

The LUMINEU and CUPID-Mo collaboration, adopted a decoupling strategy with the help of three springs on each tower hosting three to four detectors. They clearly demonstrated the benefit of this suspended structure in the low-background EDELWEISS cryostat at Modane \cite{cupidMo}.

In the experiment AMoRE, the detector array is fixed to a platform which is suspended below the mixing chamber by four phosphor bronze springs \cite{key-11}.

Other implementations of passive decoupling systems on dry cryostats with a large panel of physics applications can be found in \cite{etc,key-12}.\\

In this work, we fully characterize the use of a suspended tower based on a $\sim25$~cm long free elastic-pendulum with one stainless steel spring at a single fixing contact point. The motivation of this choice, discussed in the next section, is to reduce the number of vibration modes and to have a high attenuation along both vertical and radial directions.

This work has been done using the dry cryostat at the \emph{Institut de Physique Nucl\'eaire de Lyon }(IPNL) provided by Cryoconcept \cite{key-4}. The first two  stages (50~K and 4~K) of this DDR are thermally coupled by low-pressure gas exchangers (Hexagas\textsuperscript{TM}) to avoid any mechanical contact between the pulse-tube cold head and the dilution unit. Recent works in \cite{key-3} have demonstrated how the use of an edge-welded bellow to mechanically decouple the cold head from the cryostat frame considerably reduced vibrations and improved low-temperature performance. However, despite these improvements, some residual vibrations, espacially in the radial direction, still impact the bolometer performance, hence calling for more investigations that leads to this work.

In section~\ref{sec:Design} we describe the theoretical calculations related to an elastic pendulum and we present a design of suspended tower based on it. In section~\ref{sec:Vibration} we present measurements of the vibration levels of this newly designed suspended tower, both in terms of acceleration and displacement. Finally, in section~\ref{sec:Detector} we give a detailed comparison of cryogenic bolometer performance when mounted directly on the mixing chamber and when integrated in the suspended tower.

\section{An elastic-pendulum based suspended tower}
\label{sec:Design}
In this section, theoretical calculations for an elastic-pendulum are presented. This leads to design a prototype of the tower with a particular emphasis on its implementation in the IPNL DDR unit.

\subsection{Modeling approach of an elastic-pendulum suspended tower}
\label{sec:Design1}
The main focus is to mitigate the transmission of both vertical and radial vibrations, with a particular emphasis on the radial modes as these are less efficiently damped from our pulse-tube cold head decoupling \cite{key-3}.

The existing decoupling systems discussed in introduction are based on the principle of passive vibration mitigation using springs to hang the array of cryogenic bolometers on the vibrating dilution unit. The following approach adopts a similar strategy, except that the main aim of the system is to mitigate the radial vibrations which were found to be the dominant ones in our work. To do so, the whole detector assembly is suspended by a long pendulum composed by a single free spring. It is designed in such a way it minimizes the coupling to any transverse modes.\footnote{In the case of a plate suspended by several springs, the system can pitch.}

\begin{figure}[htbp]
\centering 
\subfloat[]{\label{fig:edge-a}\includegraphics[width=.36\textwidth]{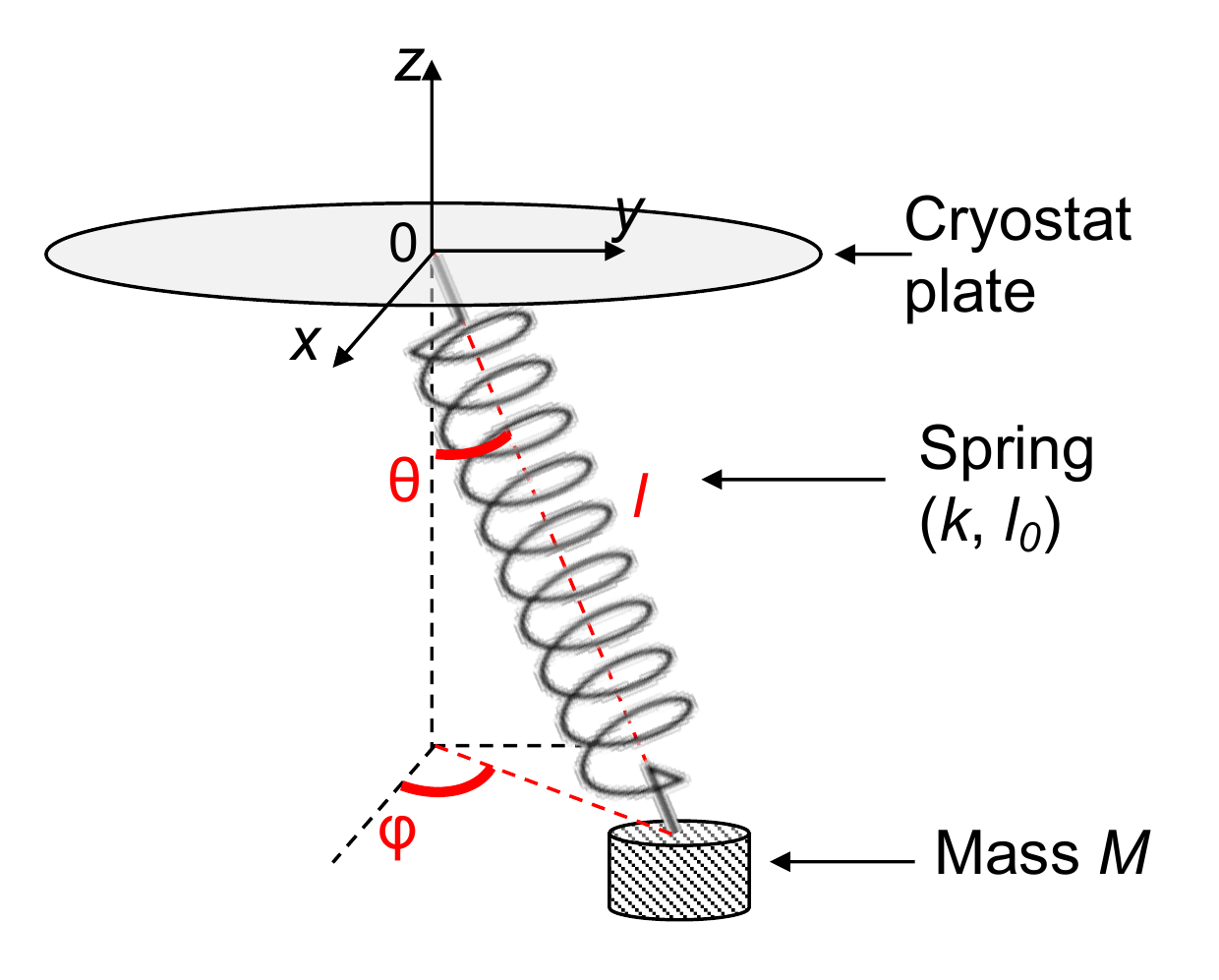}}
\qquad
\subfloat[]{\label{fig:edge-b}\includegraphics[width=.36\textwidth]{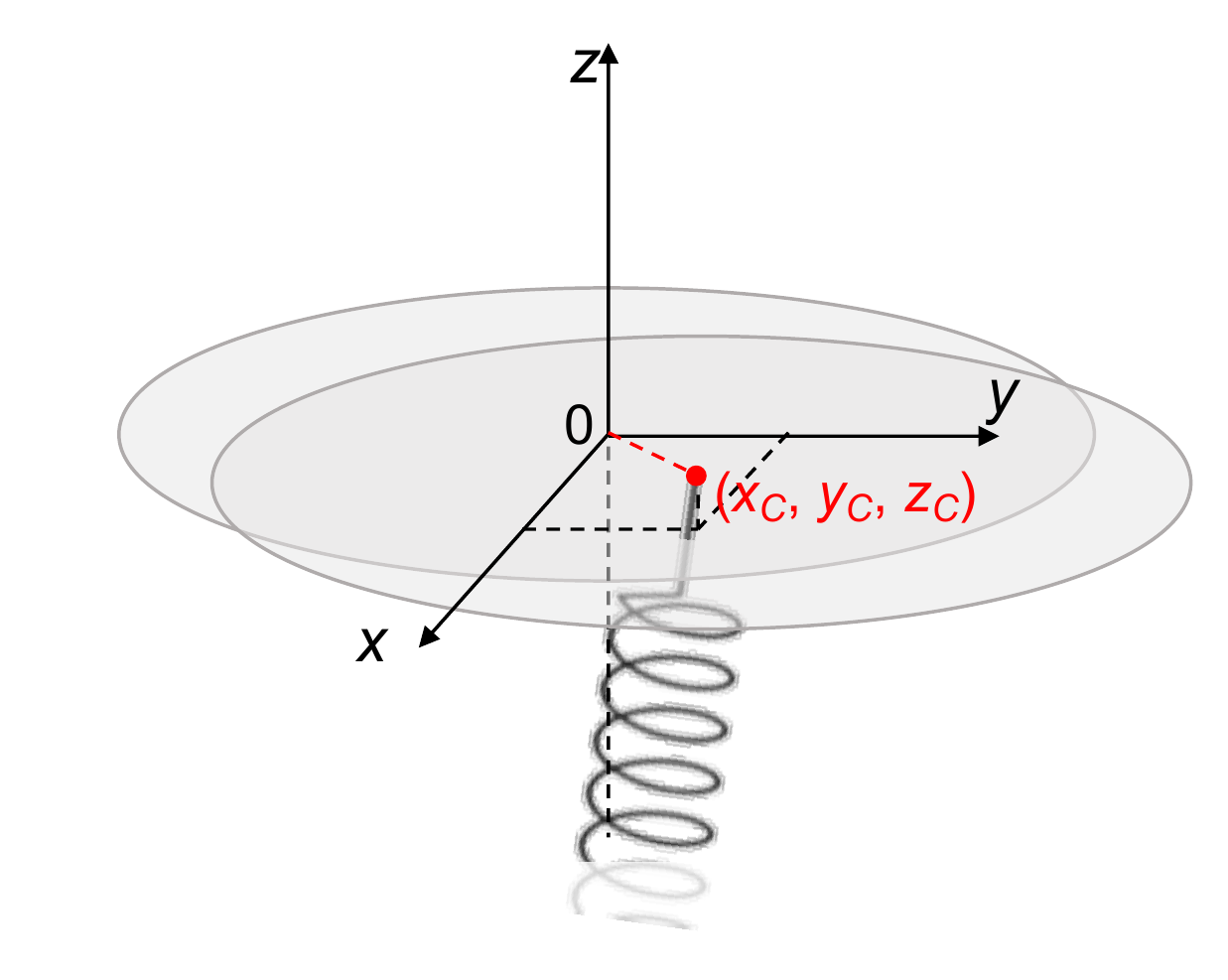}}
\caption{Schematic description of an elastic pendulum (a) which motion is induced by the perturbations of the cryostat plate (b). The coordinates $(x_{C},y_{C},z_{C})$ describe the displacement of the cryostat plate $C$ from its origin \emph{0}. \label{fig:Theoretical-pendulum}}
\end{figure}

The proposed solution is to mount the detectors onto an elastic pendulum as described in Figure~\ref{fig:Theoretical-pendulum}. A mass $M$, representing the array of detectors, is suspended at one stage of the cryostat via a single spring with rest length $l_{0}$ and an elastic constant $k$. The restoring force of the spring, $\overrightarrow{F_{R}}$, is proportional to its elongation as follows: $k\cdot(l-l_{0})$, where $l$ is the length of the spring.

In absence of perturbation, the equilibrium state is reached when the restoring force of the spring  compensates the weight of the mass, $\overrightarrow{P}+\overrightarrow{F_{R}}=\overrightarrow{0}$:
\begin{equation}
-M\cdot g+k\cdot(l_{\textrm{eq}}-l_{0})=0\label{eq:Equilibre-Pendule},
\end{equation}
where $l_{\textrm{eq}}$ is the length of the spring at the equilibrium, and $g=9.81\,\textrm{m/s}^{2}$, the Earth gravitational constant.

Due to the vibrations, the displacement of the cryostat plate, characterized by the coordinates $(x_{C},y_{C},z_{C})$, induces the motion of the pendulum. It can be described by the Lagrangian formalism (as described e.g. in \cite{key-14}) as a function of the general coordinates $q_{i}=\{l,\,\theta,\,\varphi\}$, with the length $l$ of the spring, the vertical angular deviation $\theta$ of the spring, and the azimuthal angle $\varphi$ within $(xy)$ plane. The Lagrangian $L$ of this elastic pendulum is expressed as the difference between the kinetic energy $T=\frac{1}{2}\cdot M\cdot v^{2}$ and the potential energy $V=\frac{k}{2}\cdot(l-l_{0})^{2}+M\cdot g\cdot z$, with $z=z_{C}-l\cdot\cos\theta$. Then, the motion is described by the Euler-Lagrange equations derived from $\frac{\textrm{d}}{\textrm{d}t}\left(\frac{\partial L}{\partial\dot{q}}\right)-\frac{\partial L}{\partial q}=0$.

The 3-D description of the elastic-pendulum can be divided into two pseudo-independent equations if we consider the approximation of small angle \cite{key-14,key-13}. Thus, the vertical oscillations of the pendulum are dominated by the restoring force of the spring. The natural frequency for vertical modes is then given by:
\begin{equation}
f_{0,\textrm{vertical}}=\frac{1}{2\pi}\cdot\sqrt{\frac{k}{M}}\;{\equiv}\;\frac{1}{2\pi}\cdot\sqrt{\frac{g}{(l_{\textrm{eq}}-l_{0})}}\label{eq:Freq-Vertical}.
\end{equation}
\textbf{}
The second equation describes the radial oscillations of the pendulum. Here, they are dominated by the length of the pendulum which can be considered as $l_{\textrm{eq}}$. In this way, the natural frequency for radial modes is given by:
\begin{equation}
f_{0,\textrm{ radial}}=\frac{1}{2\pi}\cdot\sqrt{\frac{g}{l_{\textrm{eq}}}}\label{eq:Freq-Radial}.
\end{equation}
However, a long pendulum can be constrained by the geometry of the cryostat (e.g. in section~\ref{sec:Design2}). In the approximation of small angles, an additional wire can be used between the cryostat plate and the spring. This will not impact the vertical behavior. Hence, $l_{\textrm{eq}}$ in Eq.~\ref{eq:Freq-Radial} can be replaced by the new total pendulum length given by $l_{\textrm{p}}=l_{\textrm{wire}}+l_{\textrm{eq}}$.

From these approximations, we can estimate the theoretical resonance frequency of the elastic pendulum depending on the spring constant, the pendulum length and the total mass of the detector assembly. Interestingly, one can notice that as $l_{\textrm{p}} \geq (l_{\textrm{eq}}-l_{0})$, the natural frequency in the radial direction is necessarily lower than in the vertical direction.\\

As already mentioned,  thanks to the use of a single spring holding system, we avoid any transverse momentum related natural frequencies which could populate the vibration spectrum at high frequencies. Therefore, in this case, we expect the suspended tower to have transfer function response to the pulse-tube vibrations under the form of a 2\textsuperscript{nd} order low-pass filter with a single resonance frequency in both directions $i=\left\{ \textrm{vertical, radial}\right\}$:
\begin{equation}
H(\omega_{i})=\frac{\omega_{0,i}^{2}}{\omega_{0,i}^{2}-\omega_{i}^{2}}\label{eq:Transfer-function},
\end{equation}
where $\omega_{0,i}$ is the natural pulsation of the suspended tower and $\omega_{i}$ is the pulsation induced by the cryostat ($\omega_{i}=2\pi\cdot f_{i}$).

\subsection{The suspended tower}
\label{sec:Design2}
The Hex UQT dry cryostat at IPNL is used for the \emph{R\&D} activities in order to design and to test next generation detectors for the EDELWEISS experiment dedicated to the direct detection of dark matter \cite{key-5,OptiEDW}.\footnote{The dry cryostat at IPNL is based on the Hexadry Standard model (Hex Std) upgraded to the Hexadry Ultra Quiet Technology (Hex UQT) one \cite{key-3}.} Two main arguments motivate a low-pass filter suspended tower design: 1) the heat signal bandwidth of the tested detectors has a roll-off frequency around $\sim40\,\textrm{Hz}$, and 2) the natural frequency of the pulse-tube cryocooler is at 1.4~Hz. Therefore, the natural frequencies of the suspended tower in both vertical and radial directions have been tuned to be as low as possible in order to attenuate all vibrations above 1.4~Hz. According to Eq.~\ref{eq:Freq-Vertical}, a vertical resonance frequency of 1~Hz requires a spring elongation of at least $(l_{\textrm{eq}}-l_{0})\geq25\,\textrm{cm}$ and thus a total pendulum length $l_{\textrm{p}}>(l_{0}+l_{\textrm{wire}}+25)\,\textrm{cm}$. Following Eq.~\ref{eq:Freq-Radial}, a total pendulum length of at least 25~cm results in a radial frequency below 1~Hz.

\begin{figure}[htbp]
\centering 
\subfloat[]{\label{fig:edge-a}\includegraphics[width=.4\textwidth]{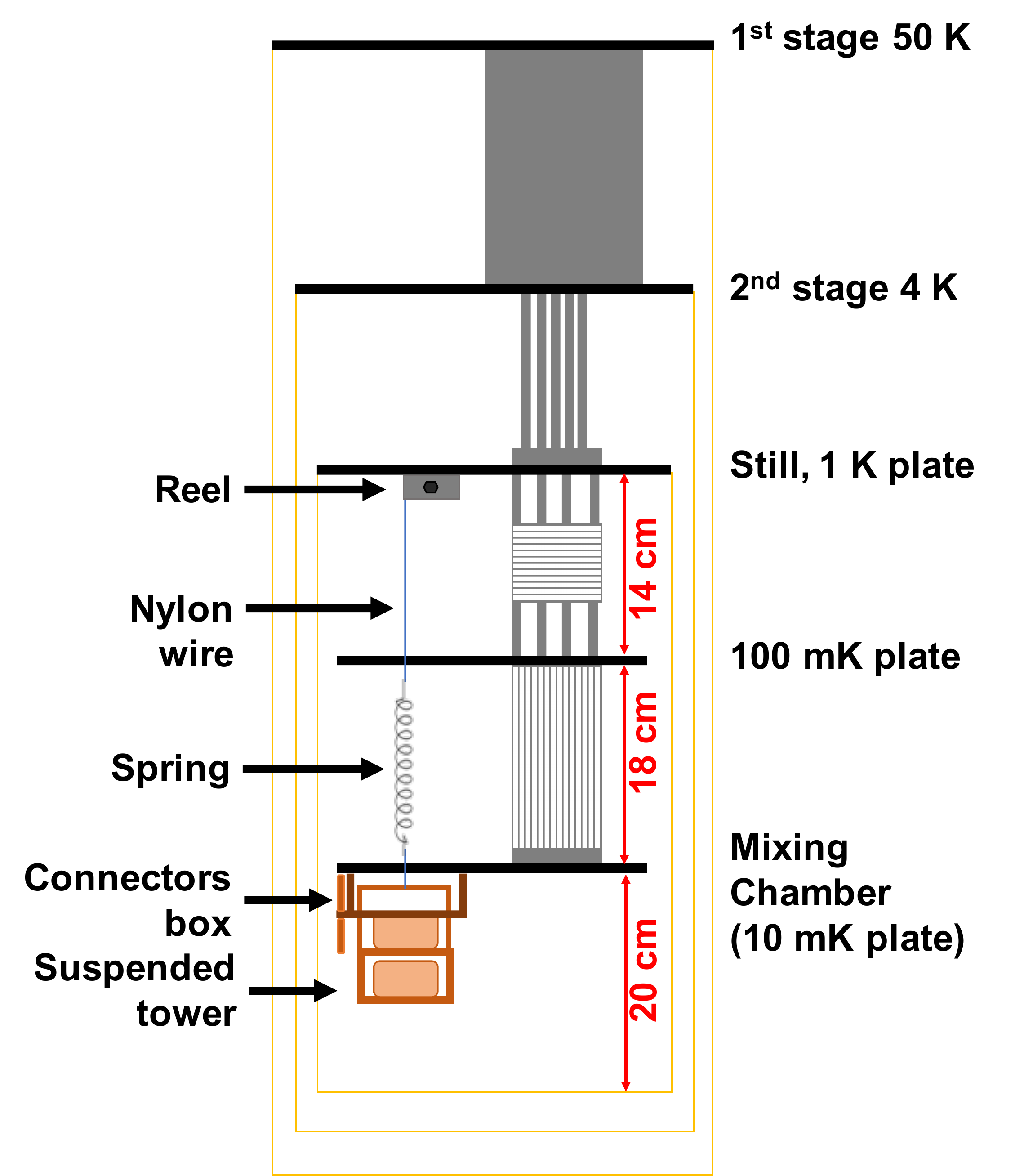}}
\hspace{8.pt}
\subfloat[]{\label{fig:edge-b}\includegraphics[width=.57\textwidth]{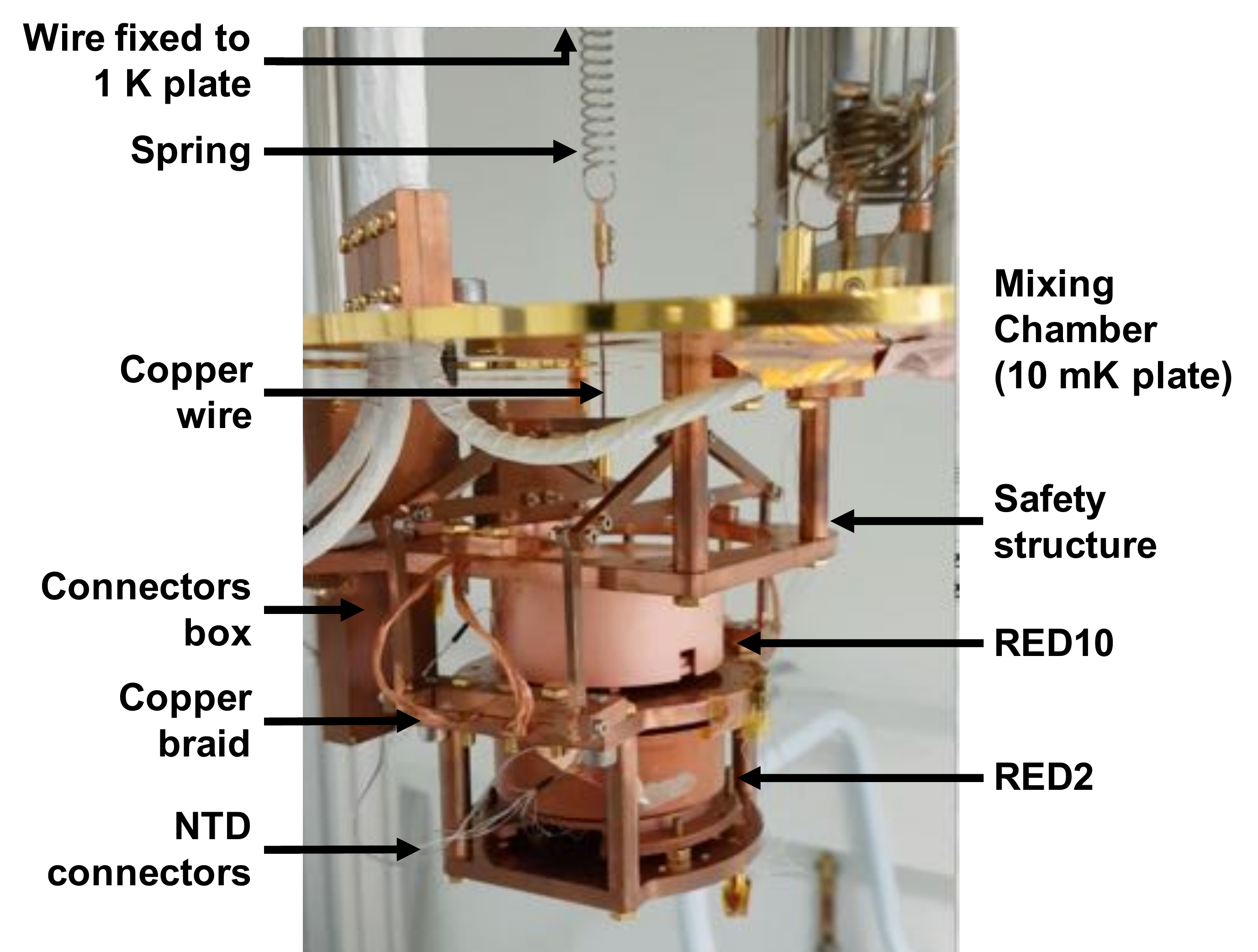}}
\caption{(a) Sketch of the set-up called "nylon wire plus spring", using a nylon wire ($l\approx19$~cm) and a stainless steel spring ($k=240$~N/m, $l_{0}=3.4$~cm) in the cryostat (thermal screens in yellow). (b) Prototype of the suspended tower installed at the dry cryostat at IPNL. Two detectors (RED2 and RED10) are mounted on it.  \label{fig:Suspended-Tower}}
\end{figure}

The challenge then comes from accommodating with the constraints imposed by the cryostat geometry. As the distance between the mixing chamber plate and the inner thermal screen is only of about 20~cm, the pendulum had to be attached to the still plate and not directly to the mixing chamber. As shown in Figure~\ref{fig:Suspended-Tower}-a, the holding strategy is carried out in three steps:  1) a nylon wire is fixed below the still plate (1~K) and running through the 100~mK stage ; 2) a stainless steel spring is attached to the nylon wire between the 100~mK stage and the MC ; 3) the suspended tower is connected to the spring via a copper wire through the MC plate. For the wire at the still plate, both nylon and Kevlar were used. Note that the elastic constant $k$ of the spring has to be carefully chosen, taking into account the total mass of the detector assembly, as its elongation is constrained by the $\sim$18~cm distance between the 100~mK stage and the MC. With this approach, the total pendulum length $l_{\textrm{p}}$ from the still plate to the center of mass of the detector assembly, can be equal to 25~cm.\\


The detector tower is shown in Figure~\ref{fig:Suspended-Tower}-b. This module can host two cryogenic detectors (RED2 and RED10 in the figure) and has a total height of 13~cm. During the installation, before attaching the spring, the tower is firmly hold by a copper frame screwed under the MC plate. This structure remains as a safety structure during cool-down and operation in case the wire should break. Connector boxes are placed close to the detector on the external side of the safety structure. They are used to connect the thermal sensors of the detectors to the readout electronics at the warmer stages. Both the suspended tower and the safety structure are made of clean CuC\textsubscript{2} copper. During operation, the suspended tower is floating and its thermalization is realized by four 10~cm long ultra-supple flat copper braids linking the safety structure and the tower. Sensors wiring and copper braids are accommodated to avoid constrains on the suspended tower.

A thick ($\sim$2~mm diameter) 5~cm long copper wire connects the top of the tower to the spring passing through the MC plate and ensured the thermalization of the stainless steel spring. Indeed, test with less conductive wires have shown that the steel spring remains warm enough to emit infrared radiation.

\section{Characterization of the vibration levels}
\label{sec:Vibration}
In this section, the vibration levels reached by the suspended tower installed within the IPNL DDR are presented and compared to the one reached by the mixing chamber plate.

\subsection{Vibration measurement method}
\label{sec:measuremethodology}
The vibration measurements method and apparatus used in this work is similar to the one described in our previous work \cite{key-3}. The set-up is composed of a high sensitivity seismic accelerometer from \emph{PCB Piezotronics} \cite{key-18}. We used the high sensitivity accelerometer \emph{PCB-393B05} model which has a gain of 10~V/g and an intrinsic noise limit of $\textrm{[0.5-0.07]}\,\upmu\textrm{g}/\sqrt{\textrm{Hz}}$ within $\textrm{[1-1000]}\,\textrm{Hz}$ frequency range.\footnote{The manufacturer indicates a possible 5\% uncertainty on the accelerometer sensitivity \cite{key-18}.} This accelerometer is connected to a \emph{PCB-480E09} signal conditioner with an anti-tribo-electric coaxial cable tightly fixed to the cryostat frame to avoid parasitic noise from the stress or vibrations of the cable. The connection between the inside and the outside of the cryostat is ensured thanks to leak-tight feedthrough. The data is digitized using a 16-bit National Instrument DAQ-6218 and processed off-line.

As in \cite{key-3}, all vibration measurements are performed at room temperature.\footnote{The variation between 300~K and 4~K, for stainless steel and copper, which are the main materials used for the rigid structures of the DDR units, of the elastic constant $k$ and Young's modulus $E$ values is expected to be small \cite{modulusE}.}  The accelerometer is fixed below the MC stage of the cryostat, and then below the bottom stage of the suspended tower thanks to an U-shaped workpiece allowing to mount the accelerometer along either the radial or the vertical direction. The output signal of the accelerometer is sampled at 10~kHz, well beyond the signal bandwidth of the accelerometer of 1~kHz. The vibration level is obtained from a Fast Fourier Transform (FFT) analysis using Hanning windowing over 5~s time windows. A Power Spectral Density (PSD) is calculated as a function of the frequency $f_{i}$ for each time window \emph{i}.
The vibration levels are then visualized through the mean Linear Power Spectral Density (LPSD) of acceleration as follows:
\begin{equation}
\textrm{LPSD}_{a}(f_{i})=\sqrt{\frac{1}{N}\cdot\sum_{i=1}^{N}\textrm{PSD}_{a}(f_{i})}\label{eq:LPSD},
\end{equation}
where $N$ is the total number of time windows (typically $N\approx50$). Using the calibration gain of the accelerometer, the $\textrm{LPSD}_a$ is given in {[}g/$\sqrt{\textrm{Hz}}${]}.

The LPSD of displacement can be reconstructed from the LPSD of acceleration (Eq.~\ref{eq:LPSD}) as follows:
\begin{equation}
\textrm{LPSD}_{d}(f_i)=\frac{g}{(2\pi\cdot f_i)^2}\cdot \textrm{LPSD}_{a}(f_i) \label{eq:LPSD_disp},
\end{equation}
where $g$ is the Earth gravitational constant. The $\textrm{LPSD}_d$ is given in {[}m/$\sqrt{\textrm{Hz}}${]}.

\subsection{Measurements of the suspended tower vibration levels}
\label{sec:measurevib}
Following the illustration shown in Figure~\ref{fig:Theoretical-pendulum}, two different suspension configurations were considered, both having a total tower assembly mass of about 1.8~kg:\footnote{The tower structure alone has a mass of 780~g. Detectors are made of a 250~g Germanium crystal inside a copper holder leading to a total detector  mass of $\sim 500$~g each.}
\begin{enumerate}
\item "nylon wire plus spring": a nylon wire followed by a spring with $k=240$~N/m and $l_{0}=3.4$~cm, leading to an elongation of 7.8~cm and to a total pendulum length $l_{p}$ of about 42~cm from the still plate to the center of mass of the tower. The corresponding theoretical radial and vertical natural frequencies are 0.8~Hz and 1.8~Hz respectively.
\item "Kevlar wire only": a single 34~cm Kevlar wire from the still plate down to the suspended detector tower leading to a total $l_{p}$ of about 42~cm. In this case we also expect a 0.8~Hz natural frequency along the radial direction and a very large natural frequency value along the vertical direction driven by the high rigidity of the Kevlar.
\end{enumerate}

\begin{figure}[htbp]
\centering 
\subfloat[]{\label{fig:edge-a}\includegraphics[width=1\textwidth]{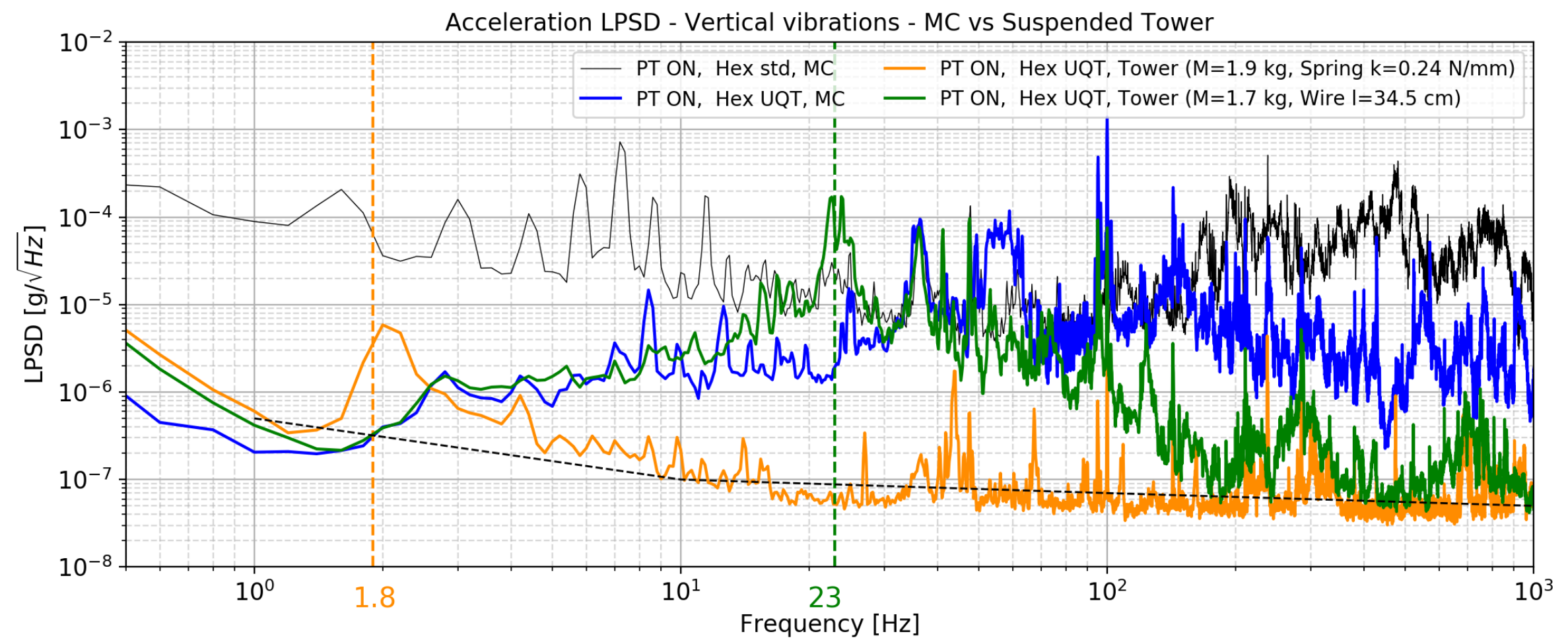}}
\qquad
\subfloat[]{\label{fig:edge-b}\includegraphics[width=1\textwidth]{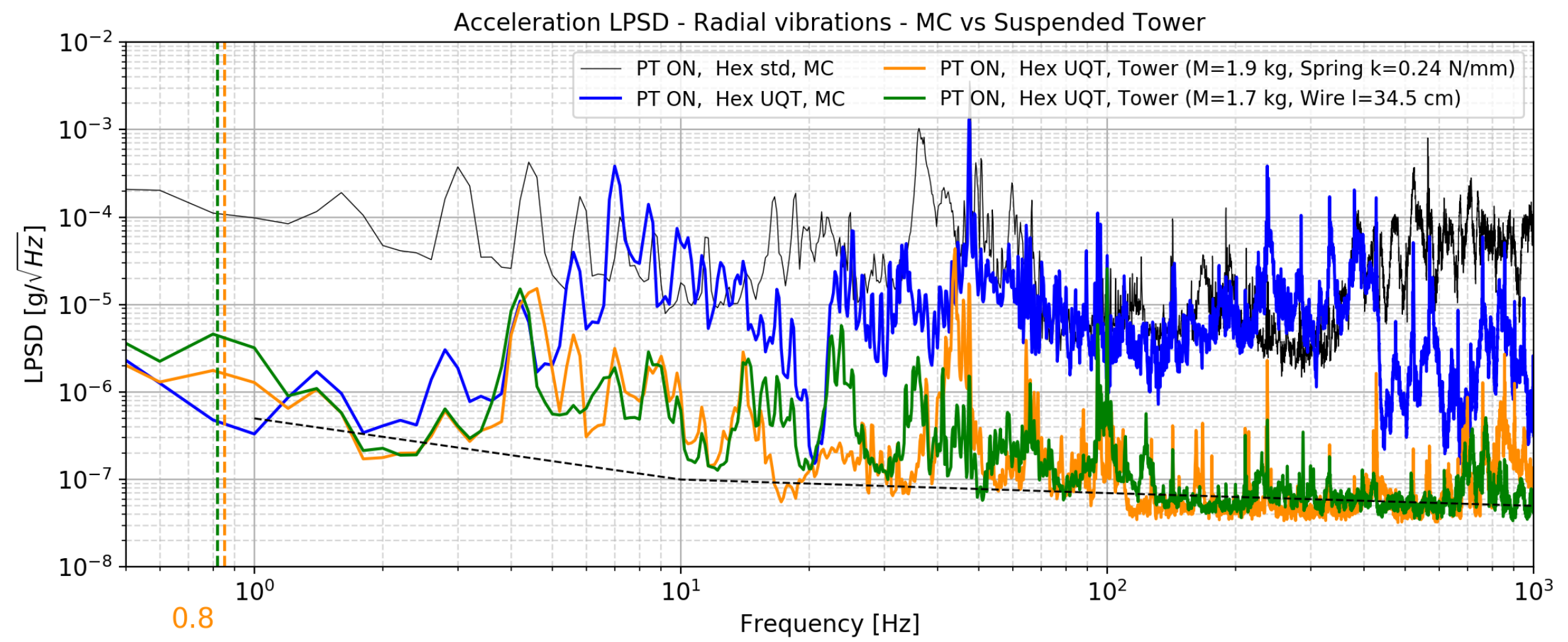}}
\caption{Measurements of the vertical (a) and radial (b) vibrations of the suspended tower with two detectors installed on it and with pulse-tube (PT) ON. Comparison of two suspension configurations. Vertical dashed lines indicate the resonance frequency of the system. The black dashed line represents the intrinsic noise of the accelerometer. Measurements within the configuration Hex Std are taken from \cite{key-3} and performed with a less sensitive accelerometer. \label{fig:Status-Tower-vibrations}}
\end{figure}

The acceleration measurements ($\textrm{LPSD}_a$) with the pulse-tube (PT) cryocooler ON are shown in Figure~\ref{fig:Status-Tower-vibrations}-a and -b for vertical and radial directions respectively. The two suspended tower configurations are shown as the green ("Kevlar wire only") and orange ("nylon wire plus spring") solid lines. The black (Hex Std) and blue (Hex UQT) solid lines correspond to the mixing chamber (MC) before and after the cold-head decoupling at 300~K realized in \cite{key-3}, respectively.\footnote{Note that even though we can clearly see the pulse-tube harmonics, the measurement Hex Std was performed with a less sensitive accelerometer.} The black dashed line is the accelerometer ultimate sensitivity taking into account all the elements of the electronic, preamplifier and acquisition readout chain.

The improvement on the vibration levels due to the suspended tower can be directly assessed by comparing the orange and green curves to the blue one (Hex UQT MC). By doing so, one can immediately notice that overall, the suspended tower drastically suppresses the induced vibrations from the surrounding environment down to a level of about 1~$\upmu\textrm{g}/\sqrt{\text{Hz}}$ to 0.04~$\upmu\textrm{g}/\sqrt{\text{Hz}}$ between 1~Hz and 1~kHz in both directions. There is an exception for suspension with only wire along the vertical direction. This improvement is particularly noticeable in the ``nylon wire plus spring'' configuration (orange solid line), where the resulting acceleration $\textrm{LPSD}_a$ is coincident with the intrinsic limit of the high sensitivity accelerometer over almost its whole bandwidth. A few additional features inherent to the vertical and radial modes are worth being mentioned in more details:
\begin{itemize}
\item Along the vertical axis (Figure~\ref{fig:Status-Tower-vibrations}-a), the $\textrm{LPSD}_a$ corresponding to the ``nylon wire plus spring'' configuration, shown as the orange solid line, clearly exhibits the expected natural frequency of about 1.8~Hz. At this frequency, the vibration amplitude is about $6\,\upmu\textrm{g}/\sqrt{\textrm{Hz}}$ and decreases down to the accelerometer ultimate sensitivity with increasing frequency. This behavior was anticipated from the transfer function \cite{key-14,key-13} of the elastic-pendulum, discussed in section~\ref{sec:Design}, which acts as a low-pass filter beyond its resonance frequency. For the second configuration with the ``Kevlar wire only'' (green solid line), the vibrations at low frequencies are clearly transmitted from the still plate, which were found to be similar to the ones at the mixing chamber. This wire-pendulum attenuates the vibrations only beyond its natural frequency, which is found to be about $\sim23\,\textrm{Hz}$  suggesting that this configuration is similar to an elastic pendulum with spring constant of about $k\approx 35000\,\textrm{N/m}$. 

\item Along the radial axis (Figure~\ref{fig:Status-Tower-vibrations}-b), the expected natural frequency of 0.8~Hz, solely related to the total pendulum length of about 42~cm, is observed in both configurations. Above this frequency, the amplitudes of the cryostat vibrations are highly reduced, again down to the impressive level of $0.1\,\upmu\textrm{g}/\sqrt{\textrm{Hz}}$ beyond 10~Hz, which is limited by the accelerometer sensitivity. Within the detector bandwidth, between 1~Hz and 40~Hz, we see that the pulse-tube natural frequency of 1.4~Hz and its related harmonics below 5~Hz are only moderately attenuated down to the $\sim\upmu\textrm{g}/\sqrt{\textrm{Hz}}$ level. We suspect that this is inherent to the shape of the transfer function of the corresponding pendulum. It is worth noticing that we repeated these measurements with and without the copper braids used for thermalization and found consistent results with what is shown in Figure~\ref{fig:Status-Tower-vibrations}-b.
\end{itemize}

As a conclusion, these results unambiguously suggest that the ``nylon wire plus spring'' configuration is the best configuration to mitigate any residual vibrations at the detector stage. In such case, we have demonstrated that the detectors will be running in an impressive sub-$\upmu\textrm{g}/\sqrt{\textrm{Hz}}$ vibration level over their signal bandwidth (1~Hz to 40~Hz) and below 0.1~$\upmu\textrm{g}/\sqrt{\textrm{Hz}}$ at higher frequencies.\\

\begin{figure}[htbp]
\centering 
\subfloat[]{\label{fig:edge-a}\includegraphics[width=1\textwidth]{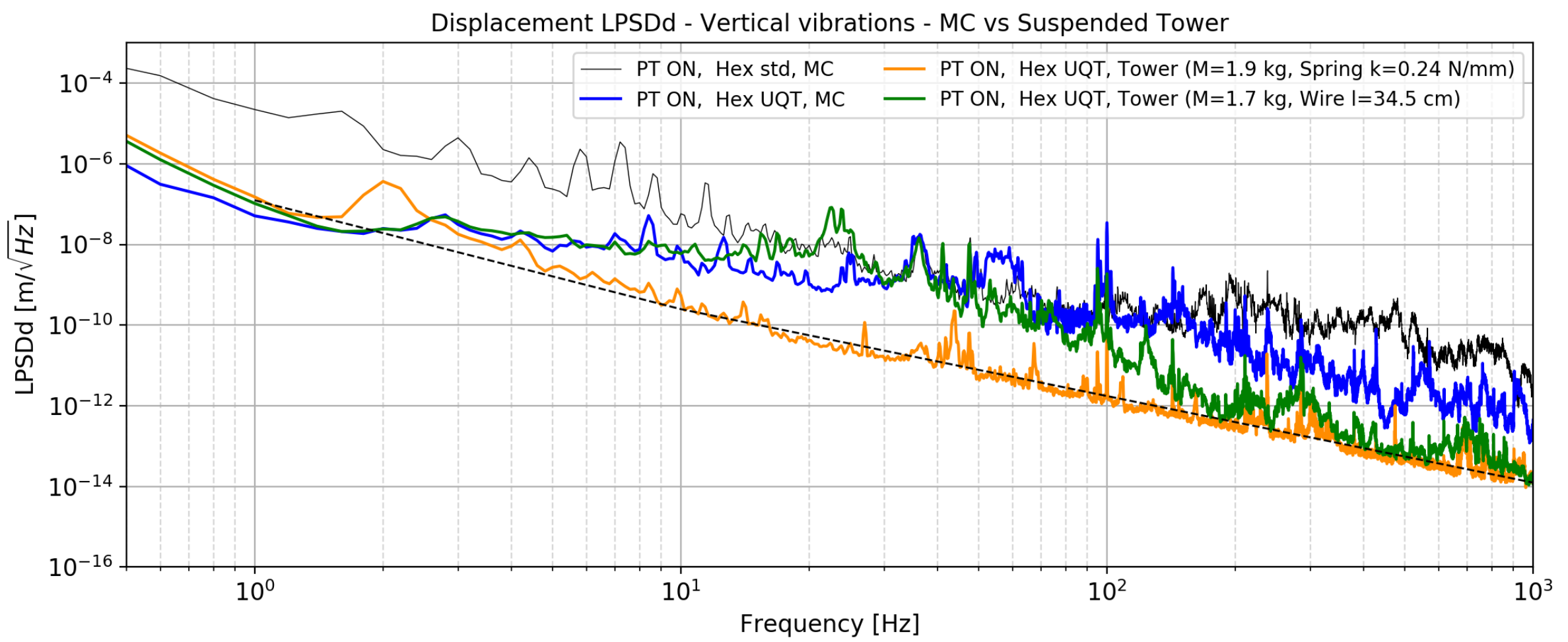}}
\qquad
\subfloat[]{\label{fig:edge-b}\includegraphics[width=1\textwidth]{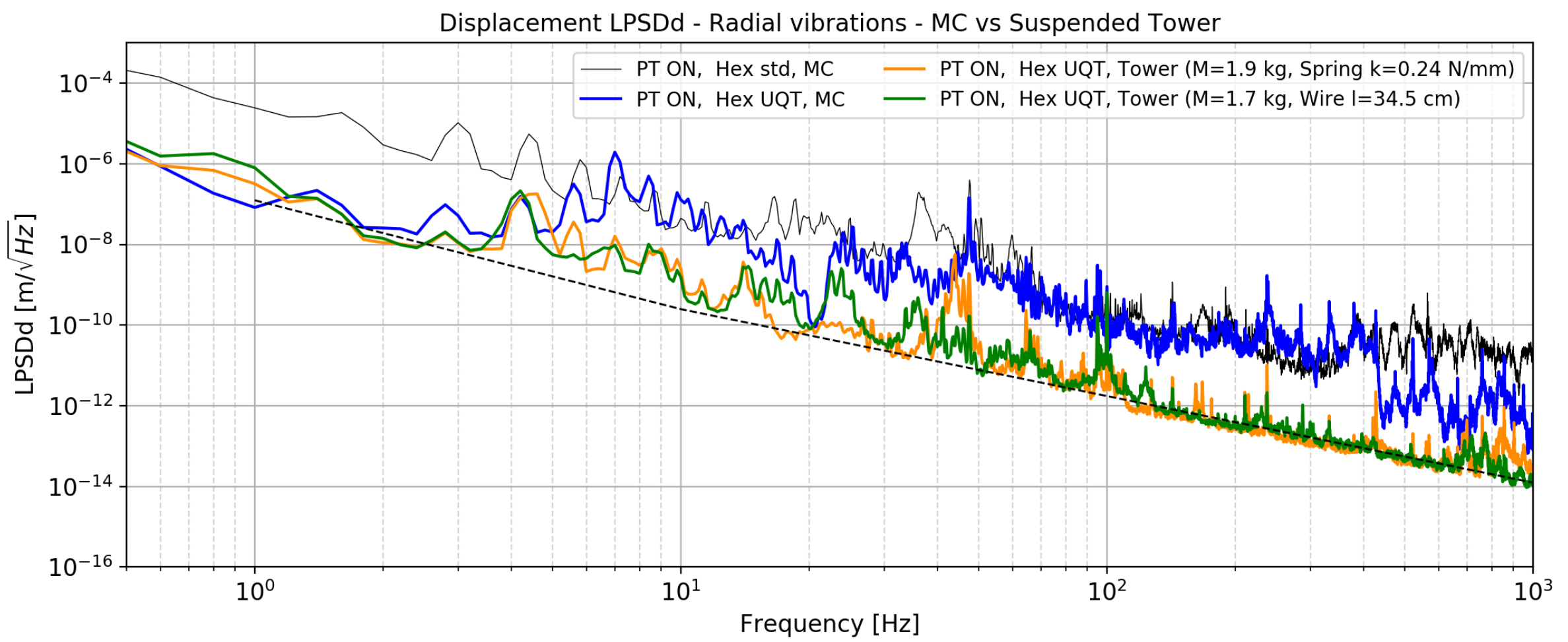}}
\caption{Measurements of the vertical (a) and radial (b) displacements of the suspended tower with two detectors installed on it and with pulse-tube (PT) ON. Comparison of two suspension configurations. The black dashed line represents the intrinsic noise of the accelerometer. \label{fig:Status-Tower-vibrations-displacement}}
\end{figure}

The vibration measurement results can also be discussed in terms of displacement using Eq.~\ref{eq:LPSD_disp}. Figure~\ref{fig:Status-Tower-vibrations-displacement} compares the displacement linear power spectral densities ($\textrm{LPSD}_d$) of the MC plate (blue for Hex UQT) and of the suspended tower along both vertical and radial directions for the two configurations under study: ``nylon wire plus spring'' (orange) and ``Kevlar wire only'' (green). Similarly to the acceleration measurements, we find that both suspension configurations drastically improve on the vibration levels which are reduced down to 1~nm/$\sqrt{\text{Hz}}$ above 5~Hz. As in Figure~\ref{fig:Status-Tower-vibrations} we notice that the first configuration, using a stainless steel spring, achieves lower displacement values at the lowest frequencies compared to the ``Kevlar wire only'' configuration.

\begin{table}[htbp]
\begin{centering}
\caption{Comparison of the RMS noise, with PT ON, of the MC plate and the suspended tower evaluated within the \emph{detector bandwidth} [1-40]~Hz and within the \emph{acoustic frequency range} $\textrm{[40-1k]}$~Hz (see details in \cite{key-3}). \label{tab:Comparison-RMS}}
\begin{tabular}{|l|r|r||r|r|}
\cline{2-5} 
\multicolumn{1}{l|}{} & \multicolumn{2}{c||}{\textbf{Acceleration}} & \multicolumn{2}{c|}{\textbf{Displacement}}\tabularnewline
\cline{2-5} 
\multicolumn{1}{l|}{} & {[}1-40{]} Hz & {[}40-1k{]} Hz & {[}1-40{]} Hz & {[}40-1k{]} Hz\tabularnewline
\hline 
\hline 
MC - Vertical & 122 $\upmu$g & 899 $\upmu$g & 0.07 $\upmu$m & 27.1 nm\tabularnewline
\hline 
MC - Radial & 254 $\upmu$g & 1002 $\upmu$g & 1.15 $\upmu$m & 91.6 nm\tabularnewline
\hline 
\hline 
Tower "nylon wire plus spring" - Vertical & 3.8 $\upmu$g & 4.5 $\upmu$g & 0.23 $\upmu$m & 0.19 nm\tabularnewline
\hline 
Tower "nylon wire plus spring" - Radial & 11.7 $\upmu$g & 34.0 $\upmu$g & 0.22 $\upmu$m & 4.2 nm\tabularnewline
\hline 
\hline 
Tower "Kevlar wire only" - Vertical & 211 $\upmu$g & 110 $\upmu$g & 0.12 $\upmu$m & 9.7 nm\tabularnewline
\hline 
Tower "Kevlar wire only" - Radial & 11.3 $\upmu$g & 15.4 $\upmu$g  & 0.39 $\upmu$m & 0.45 nm\tabularnewline
\hline 
\hline 
Intrinsic noise limit & 0.23 $\upmu$g & 0.04 $\upmu$g & 0.06 $\upmu$m & 0.07 nm\tabularnewline
\hline 
\end{tabular}
\par\end{centering}
\end{table}

An overall comparison can be done by calculating the standard deviation (RMS) in acceleration and displacement over two frequency regions: the \emph{detector bandwidth} [1-40]~Hz and the so-called \emph{acoustic frequency range} [40-1k]~Hz \cite{key-3}. The RMS values are then obtained by integrating the linear power spectral densities following:
\begin{equation}
\textrm{RMS}\mid_{f_{1}}^{f_{2}}=\sqrt{\sum_{f_1}^{f_2}(\textrm{LPSD})^{2}\cdot\Delta f}\label{eq:RMS},
\end{equation}
where $\Delta f$ is the discrete frequency step ($\Delta f$=$\frac{1}{tw}$ with $tw=5\,\textrm{s}$, the time window chosen to perform the FFT analysis) and $f_{1}$, $f_{2}$ are the limits of the frequency range. The results are shown in Table~\ref{tab:Comparison-RMS} where, for the sake of completeness, we also give  the RMS values derived from the sensitivity limitation of the accelerometer, shown as black dashed lines in Figures~\ref{fig:Status-Tower-vibrations} and ~\ref{fig:Status-Tower-vibrations-displacement}.

From Table~\ref{tab:Comparison-RMS}, one can clearly appreciate the huge gain in using a suspended tower, for both vertical and radial modes. In most cases, the RMS value is reduced by a factor of 20 to 30 in acceleration within the \emph{detector bandwidth}. One notes that the RMS value is not improved along the vertical direction for the configuration with wire alone. The RMS value in displacement is improved by a factor 3 to 5 only along the radial direction. As discussed above, this improvement is particularly impressive with the first suspension configuration "nylon wire plus spring". The only case where this configuration does slightly worse than the ``mixing chamber'' case, is in the RMS displacement within the [1-40]~Hz frequency range in the vertical mode. This degradation of a factor of 3 is due to the presence of the natural frequency peak at 1.8~Hz, heavily weighted in the RMS calculation due to the $\propto 1/f^2$ dependence of the displacement LPSD.

Finally, in Figure~{\ref{fig:Status-Tower-vibrations-PTonoff}, the impact of the pulse-tube cryocooler is observed by comparing measurements PT ON and OFF. The mixing chamber vibration levels shown as the blue (PT ON) and purple (PT OFF) solid lines are compared to the ones obtained with the suspended tower with a spring shown as the orange (PT ON) and red (PT OFF) solid lines. The main differences observed between these two PT configurations are coming from few pick-ups lines which are mostly at high frequencies and arise from acoustic perturbations originating from: e.g. the noise of the rotary valve, the audible whistling noise from the high-pressure gas going through the corrugated pipes and flex hoses. At low frequencies ($<40$~Hz), in the detector bandwidth, the impact of the PT is clearly negligible on the suspended tower. As a matter of fact, by comparing the ``mixing chamber'' PT OFF case (purple solid line) to the ``suspended tower'' PT ON case (orange solid line), one can further conclude that the spring-suspended tower is not only efficient at reducing the PT-induced noises but also most of the whole setup-related vibrations from the building, the cryostat holding structure, and so on. We can therefore conclude from this result that vibration-wise, our detectors are largely insensitive to their surrounding environment, suggesting that they should run in optimal conditions.

\begin{figure}[htbp]
\centering 
\subfloat[]{\label{fig:edge-a}\includegraphics[width=1\textwidth]{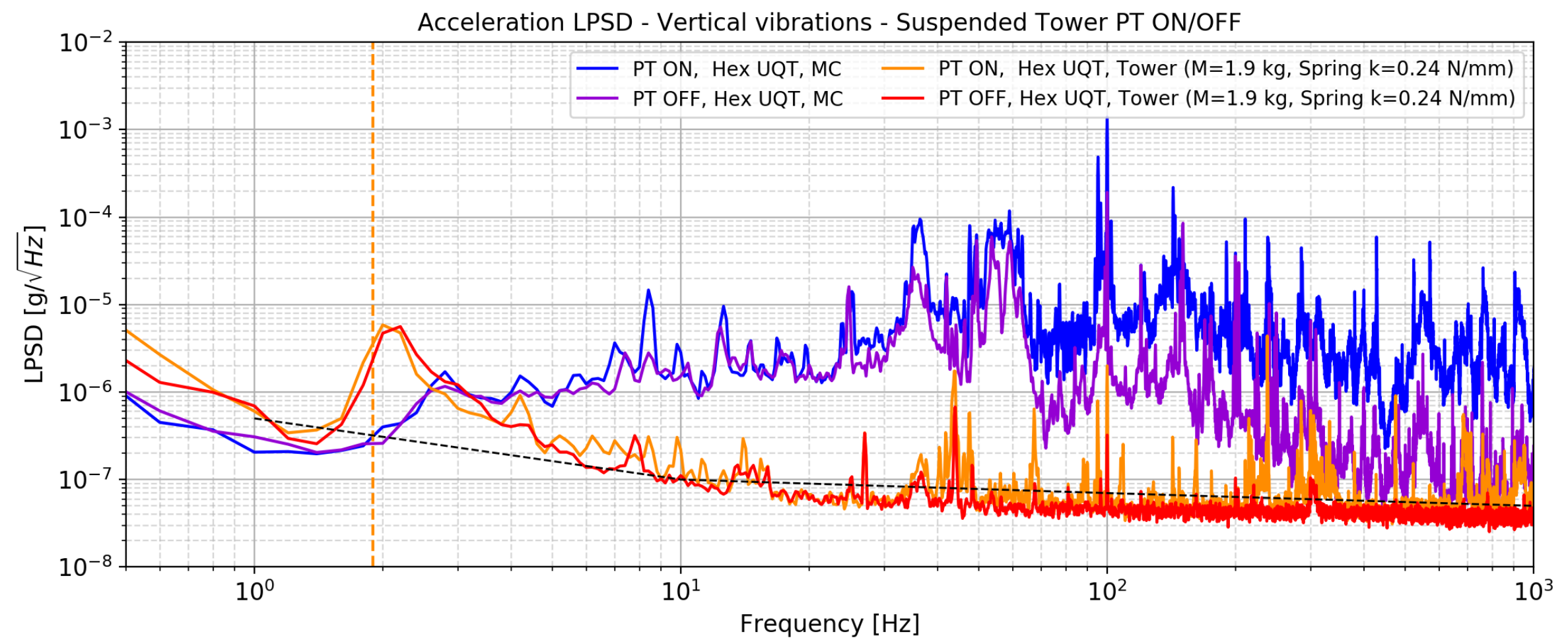}}
\qquad
\subfloat[]{\label{fig:edge-b}\includegraphics[width=1\textwidth]{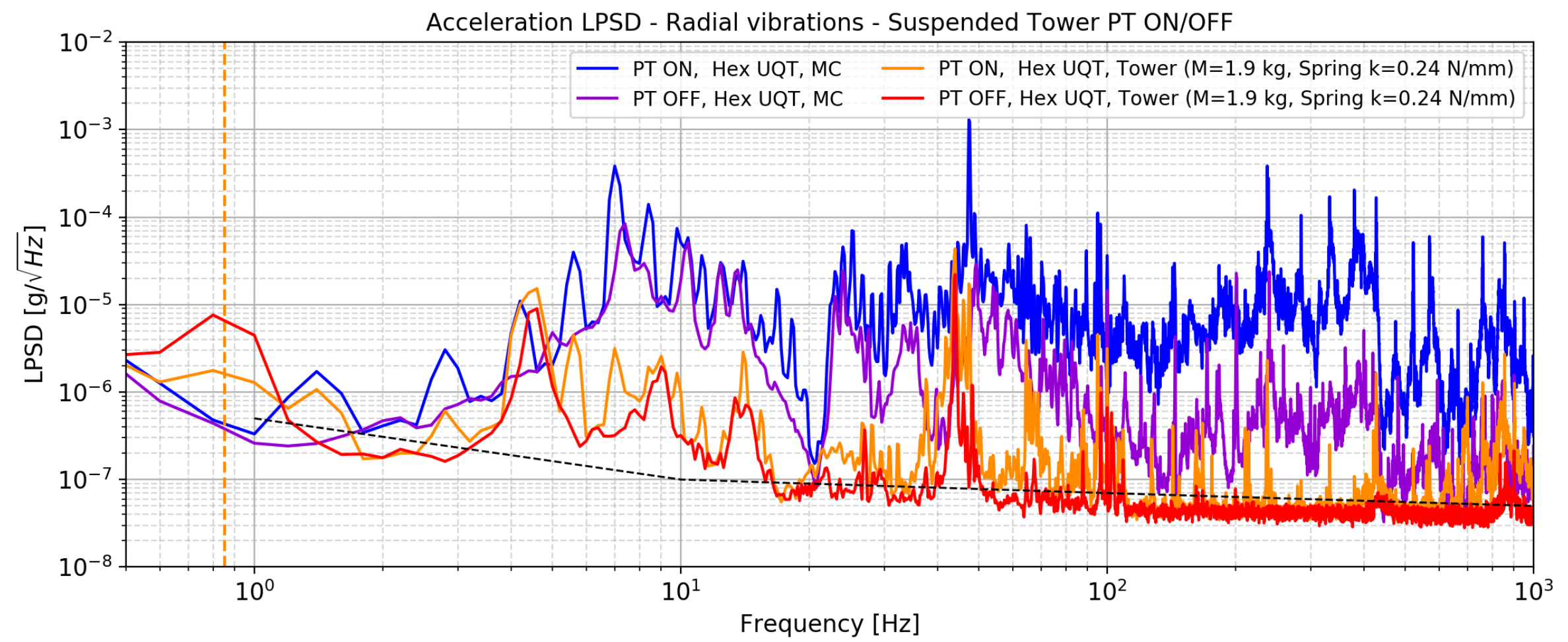}}
\caption{Comparison of the vertical (a) and radial (b) vibration levels of the MC plate and the suspended tower (``nylon wire plus spring'') in both PT ON and OFF configurations.\label{fig:Status-Tower-vibrations-PTonoff}}
\end{figure}

\section{Cryogenic detectors in the suspended tower}
\label{sec:Detector}
In this section, the noise performance of EDELWEISS-like detectors mounted in the suspended tower is investigated. These detectors consisting in germanium bolometers are dedicated to the optimization of the heat sensor design.

\subsection{Bolometer description and measurements}
\label{sec:DetectorDescription}
The heat energy deposited inside the Ge crystal is measured using Neutron Transmuted Doped germanium (Ge-NTD) thermal sensors glued on the crystal surface. The results discussed here have been obtained with two detectors mounted on the suspended tower with the spring (section~\ref{sec:Design2}).

The first one, RED2, was placed at the bottom of the suspended tower (see Figure~\ref{fig:Suspended-Tower}). This detector is a 250~g germanium cylindrical crystal fitted with two NTDs glued on the same planar surface. The thermal link between the crystal and the cryostat is ensured by one of the NTD via gold wires connected by ultrasonic bonding. The detector has sensitivity of about 100~nV/keV around 18~mK derived from its thermal design \cite{key-6}. The biasing of the two NTDs and the readout are performed with an EDELWEISS-like read-out system (described in \cite{key-5}) with a square modulated biasing current (called AC mode).

The second test detector, RED10, was mounted on the top of the suspended tower (see Figure~\ref{fig:Suspended-Tower}). It is also a 250~g germanium crystal with only one NTD glued on the planar surface. This detector has a sensitivity of about 50~nV/keV around 18~mK. The measurements are performed with a room-temperature JFET-based electronics with DC bias (a version of this electronics is described in \cite{key-7}).

The heat signals (i.e., the voltage across the NTDs) are read-out continuously during the acquisition as shown in Figure~\ref{fig:Time-Windows-RED2-MC-Tower}-a. In order to investigate the heat noise level, we have implemented a statistical method which divides the continuous stream in 1~s time windows. Time windows containing heat pulses from interacting particles (mostly gammas from the surrounding radioactivity and cosmic muons) are rejected, and the remaining time windows are used to calculate the voltage noise LPSD (see section~\ref{sec:noise}).

\begin{figure}[htbp]
\centering 
\subfloat[]{\label{fig:edge-a}\includegraphics[width=1\textwidth]{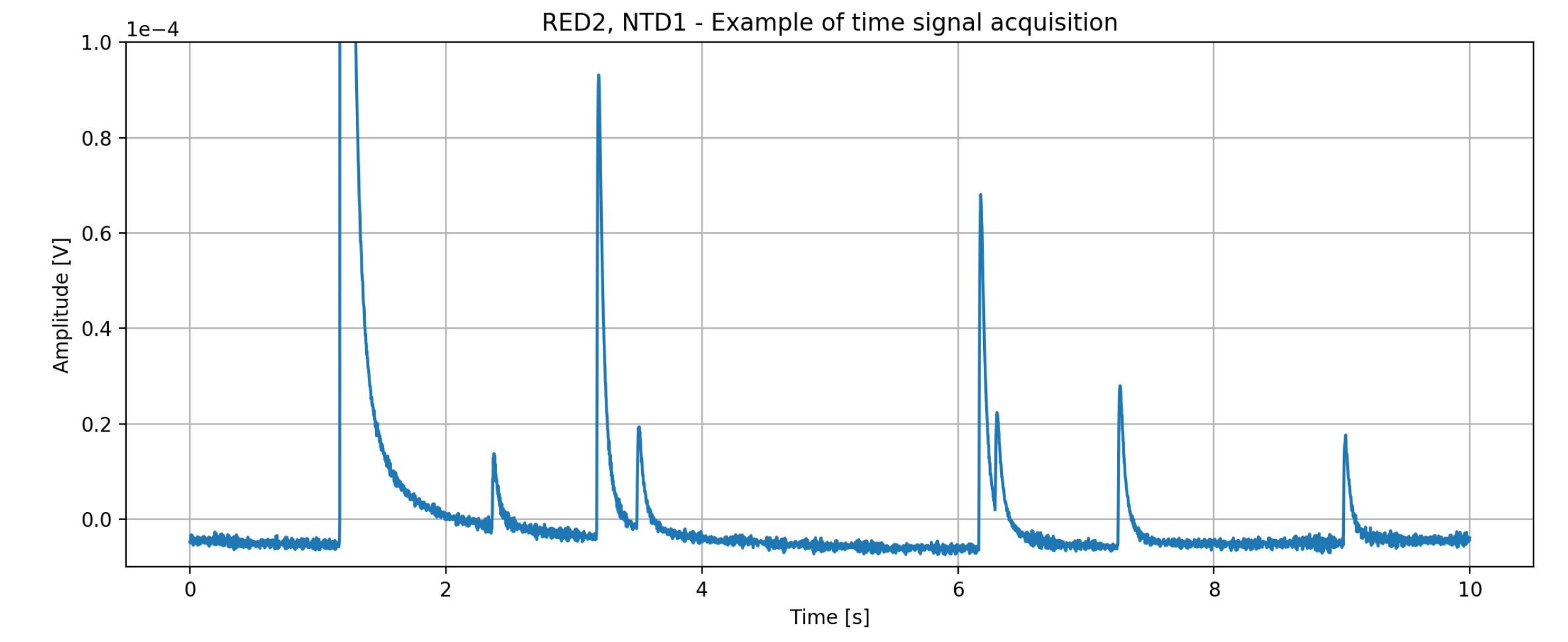}}
\qquad
\subfloat[]{\label{fig:edge-b}\includegraphics[width=1\textwidth]{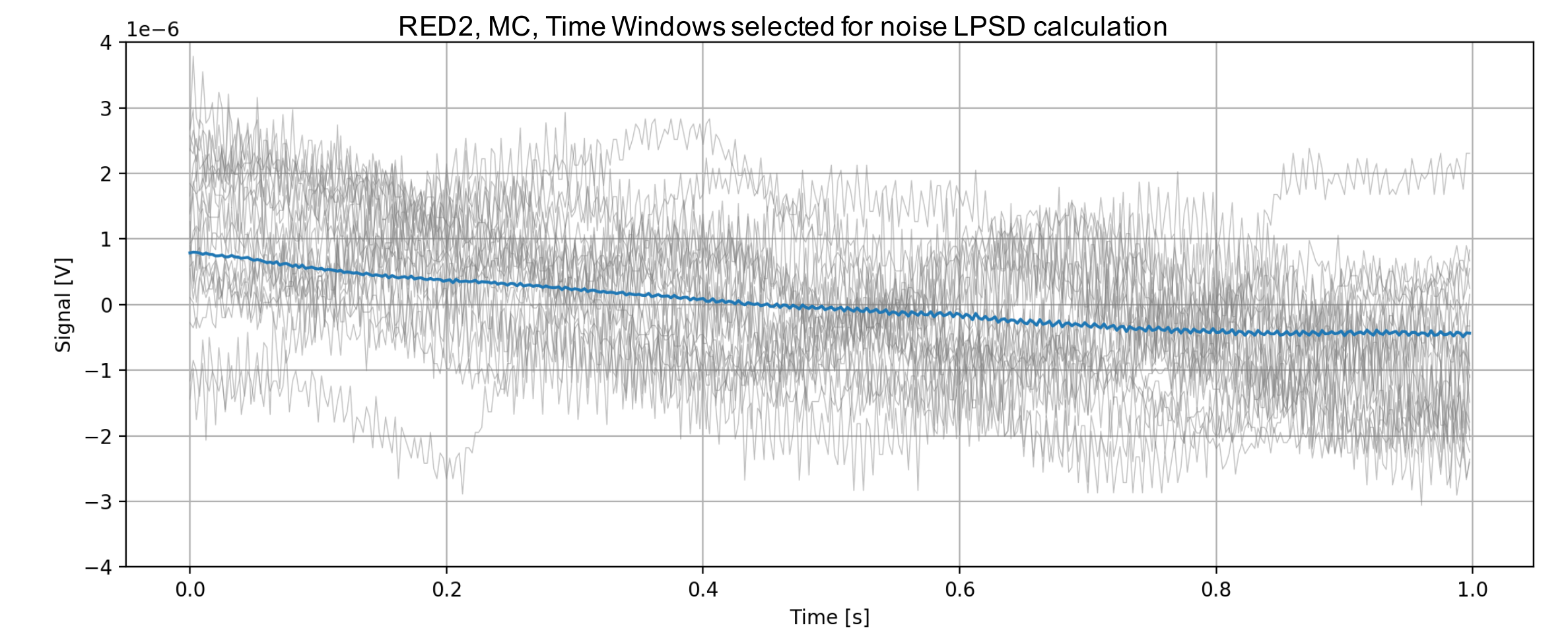}}
\qquad
\subfloat[]{\label{fig:edge-c}\includegraphics[width=1\textwidth]{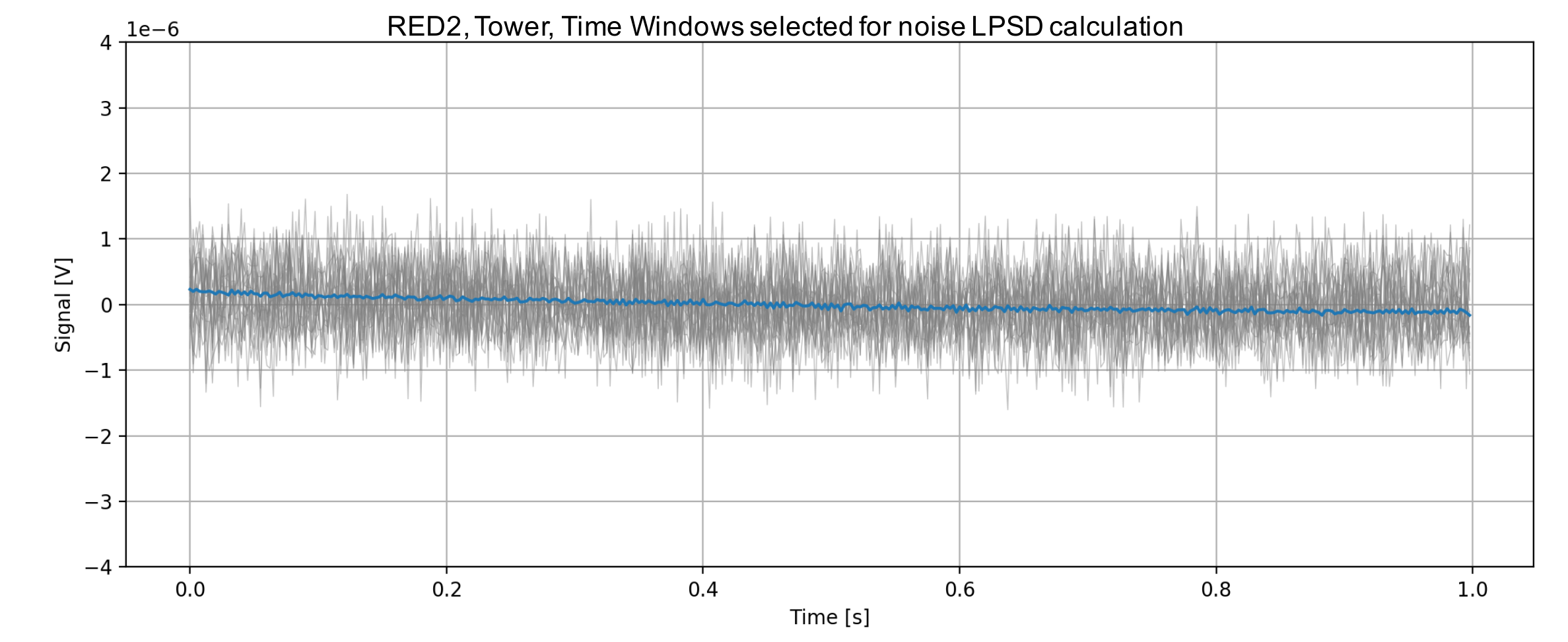}}
\caption{Example of signal for RED2 detector. (a) A 10~s signal sample showing heat pulses and noise. Example of selected 1~s time windows (gray) for the noise LPSD calculation with detector on the MC plate (b) and on the suspended tower (c). The blue line represents the average of all selected signals. Measurements performed at 18~mK and 2~nA NTD biasing current. \label{fig:Time-Windows-RED2-MC-Tower}}
\end{figure}

Figure~\ref{fig:Time-Windows-RED2-MC-Tower}-b and -c show the 1~s time windows selected for the noise LPSD calculation for one NTD of the RED2 detector. Each figure compares a few selected 1s-length time traces of the output voltage when the detector is mounted on the mixing chamber (a) and on the suspended tower (b). This comparison is made in similar conditions of biasing current (2~nA), temperature (18~mK), and with the pulse-tube cryocooler ON. As expected from the vibration measurements, when the detector is mounted on the MC plate, the baseline suffers from strong low-frequency fluctuations, compared to when it is mounted on the suspended tower. These fluctuations are due to the much higher level of vibrations on the MC. The average of the selected time windows, shown as the blue solid line in Figure~\ref{fig:Time-Windows-RED2-MC-Tower}-b and -c, has an exponential-tail time dependence due to the heat relaxation from muon interactions depositing $\sim$18~MeV in the Ge crystal at a rate of $\sim$0.5~Hz.\footnote{The minimum ionizing particle in Ge is 7.3~MeV/cm which leads to an energy peak close to 18~MeV.}  Therefore, a last selection criteria is added to the time traces used to compute the noise LPSD to reject events sitting on a muon tail. This ``muon selection'' discriminates time windows with a fitted exponential tail amplitude 2-sigma away from the baseline noise. The following noise LPSD presented in section~\ref{sec:noise} are calculated with this method.

\subsection{Bolometer noise performance}
\label{sec:noise}
In this sub-section we focus on the comparison of the bolometer noise performance between the two configurations when the detectors are mounted directly on the mixing chamber plate and when they are integrated in the suspended tower with a stainless steel spring. Figure~\ref{fig:Noise-LPSD-RED2} shows the voltage LPSDs of NTD1 (a) and NTD2 (b) from RED2 and of the single NTD of RED10 (c). The LPSDs are calculated from the selected time windows discussed above using a Hanning windowing function, and the mean LPSDs are then derived following Eq.~\ref{eq:LPSD}.

On each panel of Figure~\ref{fig:Noise-LPSD-RED2}, we show the LPSDs obtained on the mixing chamber (MC) as dashed lines, and on the suspended tower as solid lines. Also shown are the 1~V normalized heat signal bandwidth and the theoretical noise calculations based on a full electro-thermal modeling of the detectors \cite{key-6,ThermalModel}. Note that the noise expectations are different between the two detectors as using different readout electronics, AC modulation (Figure~\ref{fig:Noise-LPSD-RED2}-a and -b) and DC biased current (Figure~\ref{fig:Noise-LPSD-RED2}-c), which have different noise characteristics. When polarized, see blue (PT ON) and green (PT OFF) curves, NTDs are biased at their optimal working point: 0.5~nA for the RED2 NTDs and 4.5~nA for RED10. For completeness and to estimate the effect of the vibrations on the cabling, such as microphonics, we also show measurements with the NTDs not polarized, $I_p = 0$ (red curves).

\begin{figure}[htbp]
\centering 
\subfloat[]{\label{fig:edge-a}\includegraphics[width=1\textwidth]{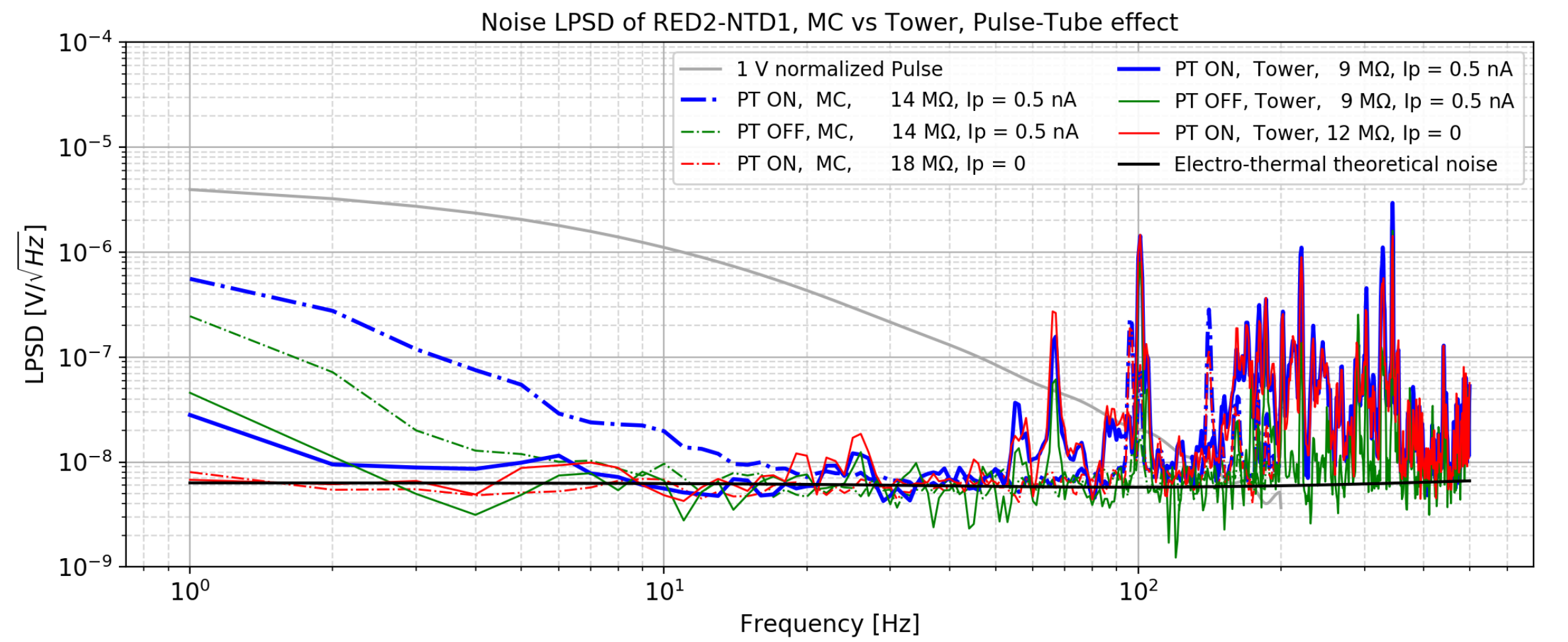}}
\qquad
\subfloat[]{\label{fig:edge-b}\includegraphics[width=1\textwidth]{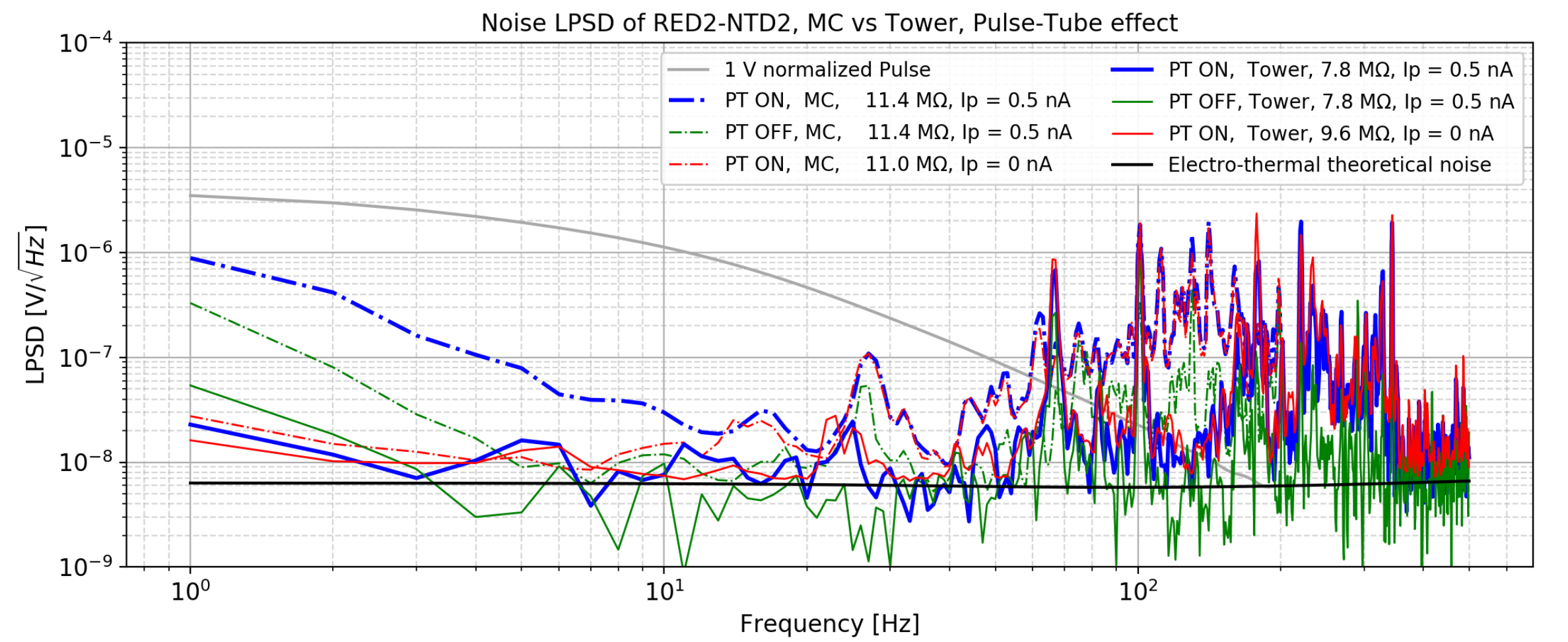}}
\qquad
\subfloat[]{\label{fig:edge-c}\includegraphics[width=1\textwidth]{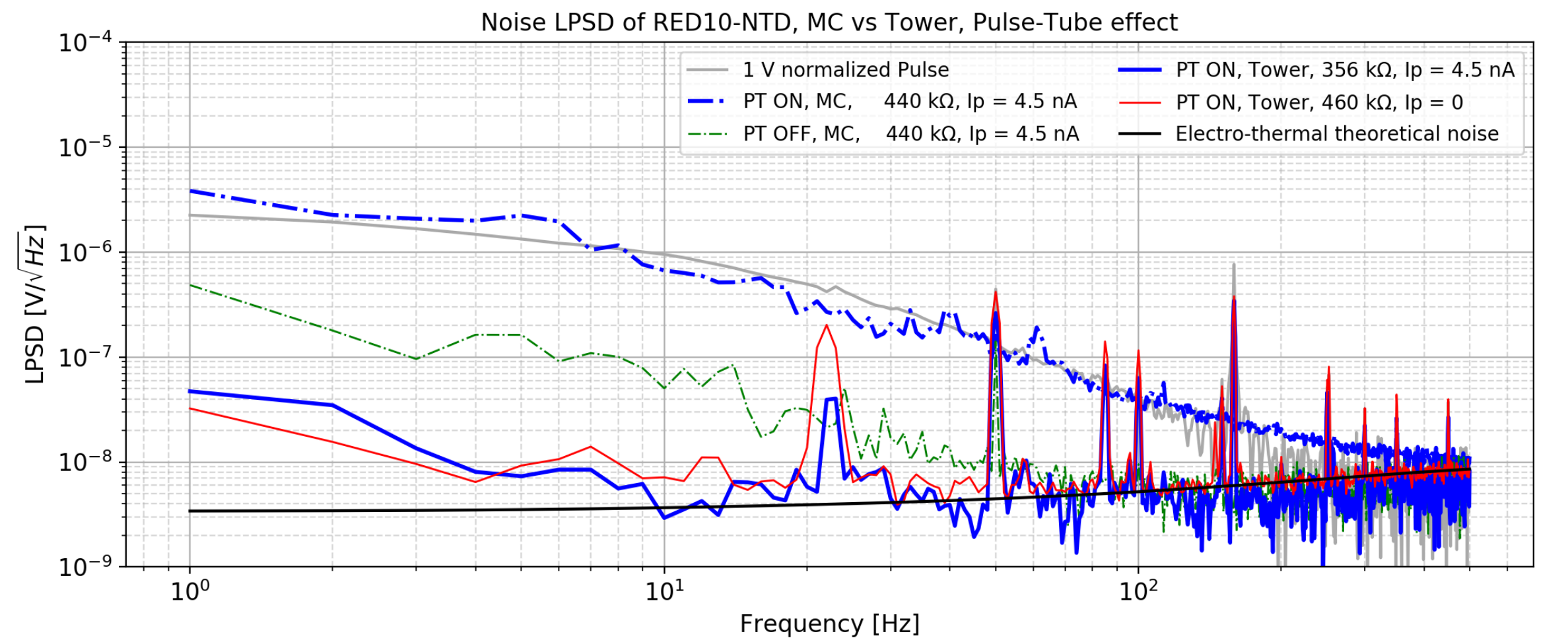}}
\caption{Noise LPSD calculated for NTD1 (a) and NTD2 (b) thermal sensors of RED2 and the NTD of RED10 (c) mounted on the MC plate (dashed line) and on the suspended tower (solid line). Measurements were performed with optimized NTDs polarization ($I_{p}$) with PT ON and PT OFF at 16~mK. The difference on the NTD resistance, between the MC and the suspended tower configurations, is due to the detector thermalization through the suspended tower. \label{fig:Noise-LPSD-RED2}}
\end{figure}

From comparing the various voltage LPSDs presented in Figure~\ref{fig:Noise-LPSD-RED2}, we can extract a few major conclusions regarding the effectiveness of the suspended tower at mitigating the vibration-induced noise on bolometers. Note that our discussion will mostly focus on the frequency range of interest, from 1~Hz to 40~Hz, as it corresponds to the detector signal bandwidth (as illustrated from the gray solid lines). The first obvious comparison is between the cases where the detectors are optimally polarized (blue curves) and either on the mixing chamber (dashed lines) or on the suspended tower (solid lines). For each of the three NTDs presented in Figure~\ref{fig:Noise-LPSD-RED2}, the LPSDs at the lowest frequencies are reduced by one to two orders of magnitude. Quantitatively, at 1~Hz with PT ON, the noise levels reach levels of $26\,\textrm{nV/}\sqrt{\textrm{Hz}}$ for NTD1 and $21\,\textrm{nV/}\sqrt{\textrm{Hz}}$ for NTD2 of RED2, and about $30\,\textrm{nV/}\sqrt{\textrm{Hz}}$ for RED10. Note that this is particularly impressive in the case of RED2 when considering the very large impedance of the NTDs $\sim$10~M$\Omega$, making them very sensitive to microphonics-induced noise. Furthermore, we observe that the noise levels obtained PT ON and with the NTD optimally polarized (blue solid lines) are almost identical to the case where the NTDs are not polarized and the pulse-tube OFF (red dashed lines), suggesting that we are limited by the electronics and not by the vibrations when the detectors are mounted on the tower. This is confirmed by the fact that the resulting noise levels are very close to the theoretical expectations illustrated by the black solid lines which do not include a $1/f$ component for neither the AC nor DC electronic simulations.\\

Interestingly, we do not see on any of the three NTDs a 1.8~Hz pick-up noise as was potentially suggested from the vertical acceleration measurements shown as the red (PT OFF) and orange (PT ON) solid lines from Figure~\ref{fig:Status-Tower-vibrations-PTonoff}. This is a major observation as it suggests that such frequency vibrations within the detector bandwidth do not limit the detector performance.\\

Finally, one can derive from Figure~\ref{fig:Noise-LPSD-RED2} that the noise LPSDs obtained with the NTDs optimally polarized on the suspended tower with PT ON (blue solid lines) are below the ones obtained with the detectors running on the mixing chamber with the PT OFF (green dashed lines). This major results indicates that not only the suspended tower damps very efficiently the vibration-induced noise from the pulse-tube cryocooler, but makes the detectors insensitive to any residual vibrations from the surrounding environment of the experiment: building, pumps. Note that this conclusion was already suggested from Figure~\ref{fig:Status-Tower-vibrations-PTonoff}.

\begin{table}[htbp]
\begin{centering}
\caption{Results of the energy resolution obtained for the RED2 and RED10 detectors. Comparison for measurements on the MC plate and on the suspended tower for the same conditions. The uncertainty on the measurement is about 5\% and is dominated by the precision of the estimation of the muon peak. \label{tab:Energy-resolution}}
\begin{tabular}{|c|c|c|c|c|c|}
\hline 
 & \multirow{2}{*}{\emph{T }{[}mK{]}} & \multirow{2}{*}{$I_{p}$ {[}nA{]}} & \multicolumn{2}{c|}{Resolution {[}keV{]}} & \multirow{2}{*}{Gain}\tabularnewline
\cline{4-5} 
 &  &  & MC plate & Suspended Tower & \tabularnewline
\hline 
\hline 
 & 18 & 0.5 & 1.5 & 0.88 & 1.7\tabularnewline
\cline{2-6} 
 & 18 & 1 & 1.2 & 0.61 & 1.9\tabularnewline
\cline{2-6} 
RED2 & 18 & 2 & 0.97 & 0.40 & 2.4\tabularnewline
\cline{2-6} 
NTD1 & 20 & 2 & 0.99 & 0.40 & 2.5\tabularnewline
\cline{2-6} 
 & 20 & 4 & 1.1 & 0.64 & 1.8\tabularnewline
\cline{2-6} 
 & 22 & 4 & 4 & 1.2 & 0.52\tabularnewline
\hline 
\hline 
 & 18 & 0.5 & 1.7 & 0.59 & 2.9\tabularnewline
\cline{2-6} 
 & 18 & 1 & 1.6 & 0.50 & 3.1\tabularnewline
\cline{2-6} 
RED2 & 18 & 2 & 1.7 & 0.37 & 4.5\tabularnewline
\cline{2-6} 
NTD2 & 20 & 2 & 1.4 & 0.29 & 4.8\tabularnewline
\cline{2-6} 
 & 20 & 4 & 1.5 & 0.31 & 4.7\tabularnewline
\cline{2-6} 
 & 22 & 4 & 1.6 & 0.25 & 4.6\tabularnewline
\hline 
\hline 
RED10 & 16 & 4.5 & 14 & 0.4 & 37\tabularnewline
\hline 
\end{tabular}
\par\end{centering}
\end{table}

\subsection{Energy resolution improvements}
\label{sec:reso}
In dark matter search, the detector performance are ultimately characterized by their baseline energy resolutions. In this work, these resolutions were calculated using a standard optimal filtering algorithm to process and recover the amplitudes of the observed events. The detector calibration is based on the position of the so-called muon peak (about 18~MeV in this detector) in the amplitude spectrum. The electro-thermal modeling from \cite{ThermalModel} confirms that this high energy calibration slightly (and conservatively) under-estimates the detector sensitivities by 5 to 10\% because of non-linearities in their thermal response above $\sim10$~MeV.\\

Table~\ref{tab:Energy-resolution} summarizes the observed energy resolutions obtained from one hour measurements with various temperature and bias current conditions, for both RED2 and RED10,  when the detectors are mounted on the mixing chamber plate or on the suspended tower. These results show a large improvement of the energy resolution for the detectors when integrated in the spring-suspended tower. The improvement is a factor two to five for the RED2 NTDs and about 40 for RED10. All the resolutions, when the detectors are mounted on the suspended tower, are below 1~keV. The best energy resolution is achieved with the NTD2 of RED2 at 22~mK and is about 250~eV (RMS). Such resolution is at the level of the best resolutions obtained with massive (250~g) cryogenic bolometers equipped with NTD \cite{key-5}. In the case of RED10, we believe that the new copper holder design used for this detector made it particularly sensitive to vibrations, hence explaining its impressive improvement of about 40 when installed in the tower. This way, RED10 was very badly performing when mounted on the mixing chamber, with an energy resolution of $\sim$15~keV, improving to a perfectly reasonable value of 400~eV (RMS) in the tower. As a conclusion, this study clearly demonstrates the need of such passive decoupling structure to make sure that the detectors are running in optimal conditions.

\section{Conclusion}
\label{sec:conclusion}
The vibrations generated by the pulse-tube within DDR represent a non-negligible background in the thermal noise of cryogenic detectors. This work proposes a suspended tower based on a free elastic pendulum using one single fixing contact point. This configuration reduces the number of possible vibration modes and highly attenuates both vertical and radial vibrations. A prototype of suspended tower has been developed based on theoretical calculations and experimental tests taking into account the constraints from the cryostat geometry and the thermalization of the detectors. The suspension and vibration decoupling is ensured by the association of a long-wire and a stainless steel spring with an elastic constant carefully chosen considering the total tower assembly mass. The suspended tower drastically suppresses the induced vibrations from the surrounding environment down to a level of about 1~$\upmu\textrm{g/}\sqrt{\text{Hz}}$ to 0.04~$\upmu\textrm{g/}\sqrt{\text{Hz}}$ within the frequency region [1-1000]~Hz in both radial and vertical directions. Furthermore, we demonstrated that this suspended tower is also efficient at reducing most of the whole setup-related vibrations from the building and the cryostat holding structure.

The efficiency of the vibration mitigation of the suspended tower is also visible by operating EDELWEISS-like cryogenic detectors mounted in it. We demonstrate an improvement by one order of magnitude of the thermal noise for two tested 250~g germanium bolometers. The impact of the pulse-tube is no more visible and the measurements in addition to the theoretical expectations show that the noise performance are no more limited by the vibration but only by the readout electronics. Thus, the baseline energy resolution of the tested bolometers is improved by a factor 2 to 40.

Now that we are no longer limited by the vibrations, we can focus on improving the thermal design of our massive cryogenic detectors to reach sub-100~eV energy thresholds as required for low-mass dark matter searches \cite{OptiEDW} and low-energy neutrino physics via their coherent and elastic scattering on nuclei (e.g. with the RICOCHET experiment \cite{key-16}).

\acknowledgments

The installation of the dry cryostat was supported by the LabEx Lyon Institute of Origins (ANR-10-LABX- 0066) of the Universit\'e de Lyon. The help of the technical staff of IPNL is gratefully acknowledged for the interventions on the cryostat and the development of the first prototype of the suspended tower. We thank Louis Dumoulin, Stefanos Marnieros, Claudia Nones and Anastasiia Zolotarova for the RED2 and RED10 detector fabrication at the CSNSM and CEA Saclay. We also thank Corinne Augier, Jules Gascon, Andrea Giulani and Denys Poda for useful comments on this manuscript.



\end{document}